\documentclass[11pt]{article}
\usepackage{emulateapj,natbib}

\newcommand{\complAB}{24.80}

\newcommand{\Pdla}{\hbox{$P_{\rm DLA}$}}
\newcommand{\Ppdla}{\hbox{$P_{+}$}}
\newcommand{\Pmdla}{\hbox{$P_{-}$}}

\newcommand{\crossPdla}	{1.62}
\newcommand{\crossPdlaerr}{1.32}
\newcommand{\crossPddla}{1.45}
\newcommand{\crossPddlaerr}{1.35}
\newcommand{\crossPm}	{-0.20}
\newcommand{\crossPmerr}{1.26}
\newcommand{\crossPp}	{-0.24}
\newcommand{\crossPperr}{2.04}
\newcommand{\crossPl}	{-0.13}
\newcommand{\crossPlerr}{1.44}

\newcommand{\CII}{\hbox{{\rm C}\kern 0.1em{\sc ii}}}
\newcommand{\CIV}{\hbox{{\rm C}\kern 0.1em{\sc iv}}}
\newcommand{\Fe}{\hbox{{\rm Fe}}}
\newcommand{\FeI}{\hbox{{\rm Fe}\kern 0.1em{\sc i}}}
\newcommand{\FeII}{\hbox{{\rm Fe}\kern 0.1em{\sc ii}}}
\newcommand{\SiII}{\hbox{{\rm Si}\kern 0.1em{\sc ii}}}
\newcommand{\AlII}{\hbox{{\rm Al}\kern 0.1em{\sc ii}}}
\newcommand{\NiII}{\hbox{{\rm Ni}\kern 0.1em{\sc ii}}}
\newcommand{\CrII}{\hbox{{\rm Cr}\kern 0.1em{\sc ii}}}

\newcommand{\ZnII}{\hbox{{\rm Zn}\kern 0.1em{\sc ii}}}
\newcommand{\OI}{\hbox{{\rm O}\kern 0.1em{\sc i}}}
\newcommand{\MgI}{\hbox{{\rm Mg}\kern 0.1em{\sc i}}}
\newcommand{\MgII}{\hbox{{\rm Mg}\kern 0.1em{\sc ii}}}
\newcommand{\HH}{\hbox{{\rm H}}}
\newcommand{\HI}{\hbox{{\rm H}\kern 0.1em{\sc i}}}
\newcommand{\HII}{\hbox{{\rm H}\kern 0.1em{\sc ii}}}
\newcommand{\lya}{\hbox{{\rm Ly}\kern 0.1em$\alpha$}}
\newcommand{\Ly}{\hbox{{\rm Ly}\kern 0.1em$\alpha$}}
\newcommand{\Ha}{\hbox{{\rm H}\kern 0.1em$\alpha$}}

\newcommand{\msun}{M$_{\odot}$}

\newcommand{\kms}{km~s$^{-1}$}

\newcommand{\D}[2]{\frac{\partial #1}{\partial #2}}

\slugcomment{Accepted for publication in ApJ, issue 10 July 2004 }

\begin{document}

\title{The clustering of galaxies  around three $z\sim 3$ damped Ly-alpha absorbers}

\author{Nicolas Bouch\'e~\altaffilmark{1}}
\affil{Dept. of Astronomy, University of Massachusetts-Amherst,   Amherst, MA 01003 USA}
\affil{Max Planck Institut f\"ur Astrophysik, Karl-Schwarzschild-Str 1, D-85748 Garching, Germany}
\affil{European Southern Observatory, Karl-Schwarzschild-Str 2, D-85748 Garching; nbouche@eso.org}

\author{James D. Lowenthal~\altaffilmark{1}}
\affil{Five College Astronomy Dept., Smith College, Northampton, MA 01063 USA; james@earth.ast.smith.edu}
\altaffiltext{1}{Visiting Astronomer, Kitt Peak National Observatory, National Optical
    Astronomy Observatory, which is operated by the Association of
    Universities for Research in Astronomy (AURA), Inc., under cooperative
    agreement with the National Science Foundation.}

\begin{abstract}
 We present out  results on the cross-correlation of Lyman break galaxies (LBGs)
 around three damped \Ly\ absorbers (DLAs) at $z_{\rm abs}\simeq 3$ from deep
  ($\mu_{I,\rm AB}(sky)\simeq 27.6$mag arcsec$^{-2}$)   $UBVI$ KPNO 4m/MOSAIC images.
 The large area of the MOSAIC images, 0.31  $\deg^2$ or $\sim 65\times
 65h_{71}^{-1}$~Mpc co-moving 
  at redshift $z=3$, allows us to probe the clustering of LBGs on scales up to
 20~Mpc co-moving.
Our survey covers a total of 1 deg$^2$ and contains $\sim$3,000
 LBGs with photometric
 redshifts between $2.8$ and $3.5$. 
 Using the redshift likelihood distributions with $m_I$ as a prior, we selected
 LBGs  within a redshift slice of width $W_z=0.15$ (corresponding to $\sigma_z$,
 the uncertainty in photometric redshifts)  centered on the redshift of the
 absorbers. 
 Within that redshift slice, we find that the DLA-LBG cross-correlation $w_{dg}$
 is $w_{dg}=(\crossPdla\pm \crossPdlaerr) \times w_{gg}$, where $w_{gg}$ is the
 LBG auto-correlation.
 This corresponds to a correlation length of $r_o=5\pm4.5h^{-1}$ (co-moving) (or
 $r_o=7\pm 6.8h_{71}^{-1}$~Mpc).
 The cross-correlation is most significant  on scales $5-10$~Mpc. 
 Through Monte Carlo simulations, we find that $w_{dg}$ is significantly greater
 than zero at the $>95$\% level.
 In three other redshift slices that do not contain a DLA, we do not find any
 evidence of clustering.
 A larger sample will enable us to discriminate between $w_{dg}/w_{gg}$ $<1$ or
 $w_{dg}/w_{gg}$  $>1$, i.e  to test
 whether DLA halos are more or less massive than LBG halos. 
 \end{abstract}



\section{Introduction}
QSO absorption lines,   including damped Ly-$\alpha$ absorbers (DLAs), and
Lyman break galaxies (LBGs)   are currently our two major sources 
of information on high redshift galaxies.  After more than two decades of 
study, the exact nature and detailed characteristics of  damped absorbers
remain unexplained. Here, we seek to constrain the properties of DLA halos
   using LBGs \citep{steidel99} as tracers of large scale structure.
   
 DLAs contain the largest reservoir of neutral hydrogen (\HI) 
at high redshifts \citep[e.g.][]{lanzetta91,lanzetta95,ellison01b}.
They contain more neutral \HI\ than all the absorbers in the Ly-alpha forest combined. 
Morever, the amount of \HI\ in  DLAs at high redshifts  corresponds 
to the amount of \HI\ in stars today at $z=0$: \citet{ellison01b} find  $\Omega_{HI}(z=3)=10^{-2.6}$,
while \citet{bell03} measure  $\Omega_*(z=0)=10^{-2.56}$ (both numbers are for $h=0.65$).
These facts led \citet{wolfe86} to put forward the `disk hypothesis', namely
that DLAs are large thick gaseous disk galaxies.  
Despite the numerous observations directed at DLAs in the past  decade
\citep[e.g. imaging studies such as][]{moller93,lowenthal95,
steidel94,steidel95,lebrun97,bunker99,fynbo99,kulkarni00,pettini00,rao00,bouche01,moller02},
this hypothesis has been debated and
the role of DLAs in galaxy formation is still not understood \citep[see the more exhaustive summary of][]{pettini03}.

Hints to the nature of DLAs are given by numerical \citep{katz96,haehnelt00,gardner01,nagamine03} and semi-analytical
\citep{kauffmann96,mo99,okoshi03} simulations of galaxy formation  at high-redshifts.  All, these simulations
 indicate that  DLAs are in majority faint (sub-L$^*$) in small dark matter halos $V_c\ll100$\kms.
Based on cross-section arguments, \citet{fynbo99}, \citet{moller02}, and \citet{schaye01} arrived to the same conclusions.
From the chemical evolution point of view, \citep{matteucci97,jimenez99,boissier03} argued
that DLAs  are  caused by gas rich low surface brightness dwarf galaxies, as seen locally in at least one case \citep{bowen01}.

However, DLAs show asymmetric profiles of their  high ionization species \citep{prochaska97,ledoux98}. 
This has been used to argue  that DLAs are, in fact, due to thick massive rotating disks \citep{wolfe86,wolfe95,prochaska97}.
But others, e.g. \citet{maller00}, \citet{mcdonald99} and \citet{haehnelt00}, showed that a large range of morphologies
can reproduce the observed kinematics: DLAs can arise 
from the combined effect of a massive central galaxy and a number of smaller satellites  or filaments.
In fact, cold gas accretion along filaments could be an important mechanism, especially at high redshifts \citep{keres03}.

Whether or not DLAs are indeed massive will lead to different  clustering properties of the galaxies around them. 
In  hierarchical galaxy formation models \citep[e.g.][]{mo96,mo02},  this clustering   yields a  
measurement of the dark matter halo mass associated with DLAs relative to that of the galaxies used 
as tracers of the large scale structure. 
In particular, if the galaxies are less (more)  correlated with the DLAs than with themselves, this will imply
that the halos of DLAs are less (more) massive than those of the galaxies. Here, we
use $z\simeq3$ LBGs  \citep{steidel99}   as tracers of the large scale structure.

In analyses similar to that presented here,
\citet{gawiser01} found no clustering of galaxies around one single DLA at $z=4$ towards BR 0951-04, and
\citet{adelberger03} found a lack of galaxies near  four DLAs (they found two within a cylinder of
radius of $5.7h^{-1}$Mpc and depth  $W_z<0.025$  whereas $\sim 6$ were expected if the cross-correlation
is the same as the galaxy auto-correlation). They argued that this is evidence that DLAs and LBGs
``do not reside in the same parts of the universe''. It is important to note that both of these
 studies were not sensitive to scales larger than
5$h^{-1}$~Mpc because   because of the small field of view available.

Other studies, however, have pointed to an over-density of galaxies near DLAs. \citet{wolfe93} combined several studies
of \lya\ emitters around DLAs and found evidence for a correlation between emitters and DLAs at a mean redshift
$\overline z=2.6$, significant at $>99$\%. 
\citet{francis93} reported the discovery of super clustering of sub-DLAs at $z\sim 2.4$ and $z\sim2.9$:
a total of four \HI\ clouds are seen towards a QSO pair separated by 8\arcmin, each being at the same velocity.
Recent results from narrow-band imaging of the Francis \& Hewett field shows that spectroscopically confirmed
\lya\ emitters are clustered at the redshift of the strongest \HI\ cloud at $z=2.9$ ($\log N_{\HI}=20.9$) towards Q2138-4427 
\citep{fynbo03}.  \citet{RocheN_00a} identified eight
        Lyman-alpha emitting galaxies near the DLA at $z=2.3$ towards PHL 
        957 in addition to the previously discovered Coup Fourr\'e galaxy 
        \citet{lowenthal91}, implying the presence of a group, 
        filament, or proto-cluster associated with the DLA. 
 Other evidence of clustering include the work of \citet{ellison01} and \citet{dodorico02}.
 \citet{ellison01} found that the DLA at $z_{\rm abs}=3.37$ towards
 Q0201+1120 is part of a concentration
 of matter that includes at least four galaxies (including the DLA) over
 transverse scales greater than $5 h^{-1}$~Mpc.
 \citet{dodorico02} showed  that out of ten DLAs in QSO pairs, five are matching
 systems within 1000\kms.
 They concluded that this result indicates a highly significant over-density of
 strong  absorption systems over separation lengths from $\sim 1$ to $8 h^{-1}$ Mpc.

Despite numerous attempts to reproduce the properties of DLAs   \citep[e.g][]{katz96,mcdonald99,mo99,gardner01,nagamine03,okoshi03}
in numerical simulations or semi-analytical models,  there are no predictions of the clustering of galaxies around DLAs.
In \citet[][in preparation]{bouche03b}, we discuss the cross-correlation of LBGs around DLAs in Smoothed Particle
Hydrodynamical (SPH) numerical simulations. There, we find that the DLA-LBG cross-correlation is
weaker than the LBG auto-correlation by approximately 30\%, indicating that DLAs reside in less massive dark
matter halos than those of LBGs.

In \citet{bouche03}, we presented a preliminary measurement of clustering of LBGs around a DLA at $z\sim3$
towards APM 08279+5255, based on part of this data set. We detected an over-density of $\Sigma/\Sigma_g\simeq 3$ at the 95\%
 level on scales $2.5<r_\theta<5$~Mpc, implying that at least some DLAs reside in dense environments.
In this paper, we expand on that study and present results on the DLA-LBG cross-correlation in our full survey of three fields.

In section~\ref{section:data},  we present the imaging data used here, our
 completeness limits and  pre-selection of high-redshift galaxy candidates.
We used photometric redshifts to select our $z=3$ LBG candidates as discussed in section~\ref{section:sample}. 
Our clustering  analysis is shown in section~\ref{section:cluster}.
The results on the DLA-LBG cross-correlation  are shown in  section~\ref{section:results},
  and discussed in section~\ref{section:discussion}.
Section~\ref{section:conclusions} contains our conclusions.

Throughout this paper, we adopt $\Omega_M=0.3$, $\Omega_\Lambda=0.7$
and $H_o=100 h$~\kms~Mpc$^{-1}$; thus, at $z\sim3$, $1\arcsec$
corresponds to  $\sim 21.5h^{-1}$~kpc and
$1\arcmin$ to $\sim 1.29h^{-1}$~Mpc, both  {\em co-moving}. At that redshift,
$H(z)\sim 4.46 H_o$, so $\delta z=0.1$ corresponds to   
$67 h^{-1}$~Mpc in co-moving coordinates~\footnote{Using the latest cosmological
parameters from WMAP ($\Omega_M=0.268$, $\Omega_\Lambda=0.728$, $h=0.71$)
changes these numbers by $\sim$ 3\%.}.

\section{The data}
\label{section:data} 
\subsection{The DLA fields}

Given the allocated telescope night, we selected three fields for the presence of
 a DLA at $z\sim 3$ and with the additional constraint that the QSO must be at a higher redshift
than that of the DLA, i.e. $z_{\rm abs}<<z_{\rm QSO}$. 
  The redshift $z\sim3$ is ideal for selecting LBGs efficiently
  using standard photometric filters. 

Our results on the QSO APM 08279+5255 field were presented in \citet{bouche03}.
The other QSOs in our survey are PC 1233+4752 ($z_{\rm em}=4.447$), with a DLA at $z_{\rm abs}=3.499$ \citep{white93}; and
SDSS J0124+0044 ($z_{\rm em}=3.840$), with  a DLA at $z_{\rm abs}=3.077$ (C. P\'eroux, 2003, private communication).
The individual QSO and DLA properties  are listed in Table~\ref{table:fields}.

Although our   DLAs do not meet the column density threshold of $\log N_{\HI}>20.3$ often quoted, the latter
 is arbitrary and based on resolution threshold of previous-generation instruments. Furthermore,
the metallicities and the \HI\ properties of `sub-DLAs' are not different to the `strict-DLA' population \citep[e.g][]{peroux03}.
For the purpose of  this study, the hydrogen column densities are $\log N_{\HI} \geq 20.0$, which
ensure that the absorption is damped and that the gas is neutral.

\subsection{Observations}
 
The observations were carried out with the MOSAIC camera at the 
  Kitt Peak National Observatory 4-m telescope on UT 2000 February 7 and 8 (run I), and
on UT 2001 September  23--26 (run II).
 Run II was photometric, some cirrus were present during run I.
The seeing was 0.9--1\arcsec.5 for both runs.

 The wide field imager MOSAIC contains eight 2k$\times$4k thinned SITe CCD. With   0\arcsec.258 per pixel,
 it has a field of view of  $35\arcmin$ on a side. The readout noise is
$\sim 6e^{-}$~pix$^{-1}$, the dark current is negligible ($\sim 5e^{-}$ hr$^{-1}$),
and the average gain is $3 e^-$~ADU$^{-1}$. Each CCD has been thinned for
detecting U-band photons. 

\begin{figure}
\plotone{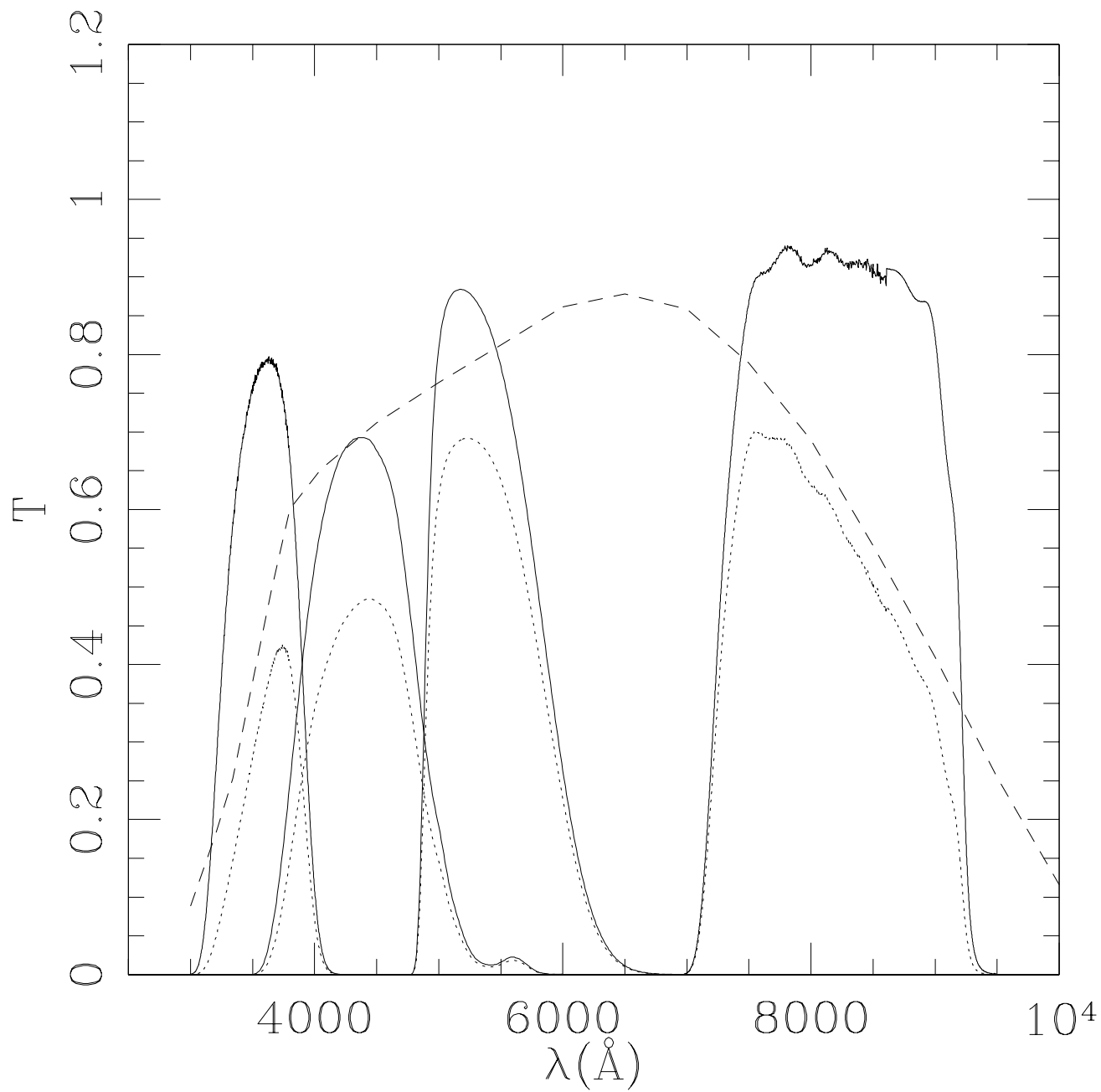}
\caption{The solid lines show the transmission curves for our four filters $U$, $B$, $V$, and $I$.
The dashed line shows the CCD response function. The dotted lines show
the filter transmission convolved with the CCD response function.}
\label{fig:filters}
\end{figure}

	We imaged our three fields through four broad-band filters---U (Stromgren) and BV\&I (Harris set)
(see Fig.~\ref{fig:filters})---
using a standard  dither pattern (five pointings) to remove cosmic rays and detector defects.
 The total integration time for each field was typically
4hr(U), 1hr(B \& V) and 4hr(I); the observations are summarized in Table~\ref{table:observations}.
In addition, we observed several  Landolt (1992) standard star fields   through each filter.

\subsection{Data reduction}

	The data were reduced with the package MSCRED (v4.1) within IRAF\footnote{IRAF is distributed
by the National Optical Astronomical Observatories, which are operated by AURA, Inc. under
contract to the NSF}(v2.11.3), following
the  reductions guidelines of the NOAO Deep Wide-Field Survey
\citep{jannuzi00}. This package was specifically designed to reduce MOSAIC data.
The reduction process contains more steps than typical optical observations and is detailed here.

We first  performed the overscan level subtraction. We then corrected for the small correction ($<$0.5\%)
 due to cross talk between adjacent chips.
 For each night's data, we removed an averaged zero frame, or bias frames, from the science images. 
 The thinned MOSAIC chips required no dark correction since the dark current
 was only 5$e^{-1}$ per hour.
 
Flat-fielding is critical to achieve precise photometry.  However, we had to deal
with two non-traditional complications: (i) the MOSAIC instrument
at the 4-m suffers from a ghost image of the pupil  in all bands
due to reflections in the optics of the camera that needs to be removed;
 and (ii) dome flat-fields may match the night sky to  only 1 or 2\% (usually larger).
Thus, sky-flatfielding in addition to dome-flatfielding was necessary and
since the pupil image is an additive light effect, it has to be removed from the flat-fields first.

In the dome flats, we  removed the pupil image   by fitting    an axially symmetric
pattern to the data themselves with the task MSCPUPIL. 
We then flat-fielded all the science frames with pupil-free dome-flats.

At this point, the pupil image was still present in the data. It is more difficult
to remove the pupil from individual science images  than from the dome-flats
because   the pupil pattern is much fainter, and   the pupil image is mixed with all the faint and
bright sources in the data. A simple threshold scheme to remove the objects is not feasible since
the pupil is still present and the data poorly flat-fielded.
Thus, to remove the pupil from the science images and to make the sky-flats, we had to extract
the pupil image from the science images themselves through the following iterative steps:
 (i) we created a sky-flat (ver. 1) from the average of the science frames with a median rejection; 
(ii) we extracted  the pupil image from the sky-flat (ver. 1)
 using the task MSCPUPIL (parameter `type' set to `data') and  removed it
 from the science frames with  RMPUPIL to
produce a `first-pass' pupil-free data;
(iii) we created another sky-flat (ver. 2) using the pupil-free data and applied it to the
data.
We found that low-level light from bright stars was creating signal in the sky-flat
even if strong min-max rejection was used.  To solve this problem, 
we created object masks by using a $2\sigma$ threshold on each of the eight CCDs
and masked out large areas around the brightest objects.
Steps (i) through (iii) were repeated using the masks and the
final sky-flat was normalized and applied to all the frames.
For the I-band only, we removed fringing using the procedure in
\citep{jannuzi00} before applying the final sky-flat.

Cosmic  ray removal was done using the task XZAP from the package DIMSUM
and customized routines. Bad-pixel masks including
 the cosmic rays and bleeding regions were constructed.

De-projecting the 8 CCDs on a single image is a two step process and requires very good astrometry.
First, using the coordinates of several hundred USNO stars, we   interactively  derived astrometric solutions
(RMS $\leq 0.5\arcsec$) with MSCCMATCH for each dithered exposure. 
Then, we mapped  the eight CCDs  onto a single image   by rebinning
the pixels to a tangent-plane projection, thus producing pixels of constant
angular size, with the task MSCIMAGE.\footnote{The parameter  `fluxconserve' was set to `no' because
the science exposures have been flat-fielded with flats equally distorted to yield a constant sky per pixel.}
This process  matches the World
Coordinate Solution (WCS) solution of all bands to a common reference position.

The individual flat-fielded, astrometrically calibrated images with a uniform zero-point
were averaged with an average sigma-clipping rejection to produce the final stacked images.
The scaling of each individual dithered image was performed interactively 
on $\sim 300$ astrometric calibration stars common to all images.

For the I-band of run II, we were not able to achieve a satisfactory
sky-flatfielding(residuals $\sim 1$\%). In order to correct for this,
  we applied a median
filtering to a block-averaged image of the stacked frame and applied the normalized 
results to the image.

   Even though all bands were deprojected to a tangent-plane solution using
the same position and orientation on the sky,
the relative pixel positions of objects in the different bands were not exactly identical
 because of dithering and effects such as flexure of the telescope, and optical distortions due to filters.
 Since it is important to have identical pixel positions for the photometry,
we had to register  and  rotate slightly ($\sim 0.003$~deg) each   image with respect to the reference
band ($U$) using $\sim 25$ stellar objects with high S/N 
in all bands throughout the field of view. 

  The  rms in the relative astrometry of $\sim 150$ stellar objects with $20<m_I<22.5$ 
  and $m_U<23.5$mag is typically  $\sim 0.4$~pix throughout the field.

\subsection{Calibration }

The standard star frames, which contained $\sim 150$ \citet{landolt92} standard stars observed
through each filter and airmasses  $1.0<X<2.5$, were reduced the same way, including the
 rebinning of the pixels to a tangent-plane projection. 
 For run I, because only a handful of standard stars were available, we  
tied the photometric solution of run I against run II on the  APM 08279+5255 field, which was re-observed in run II.

Total  fluxes of the standard stars were measured
by fitting a \citet{moffat69} profile (similar to the 7\arcsec--aperture flux used by Landolt 1992 at the 0.015 mag level).
Table~\ref{table:zeropoint} presents the best fit (computed using a linear Single Value  Decomposition algorithm)
to  the photometric    equation (e.g. Harris 1981):
\begin{equation}
m^{obs} = -2.5 \log \frac{C}{t_{exp}} + ZP + \alpha \; X + \beta \; (\rm color), \label{equation_photometry}
\end{equation}
where $C$ is the number of counts, $t_{\rm exp}$ is the effective exposure time, $ZP$ the photometric zeropoint,
$\alpha$ the airmass coefficient and $\beta$ the color term, which accounts for variations in the effective 
wavelength sampled by the filter for stars with different colors.
We achieved a flux limit of typically (m$_{AB}$) 25.5mag ($3\sigma$) in a $2\times$FWHM diameter aperture
(see  Table~\ref{table:depth} for all the flux limits).

\subsection{Source detection} 

All sources were detected in the I-band using the SExtractor (v2.1.6) software~\citep{bertin96}.
This algorithm performs source detection and photometry \citep[see][for a  summary of this package]{simard02}.
We optimized the configuration parameters to ensure
the faintest sources were detected and to optimize our  completeness.
We used 
the local background estimated in a 24-pixel wide annulus.
The images were convolved with a 2 pixel FWHM Gaussian kernel before source detection.
The detection threshold was set to 1.5 sigma with a minimum area of 5 pixels.
Bad-pixel masks are used as flag images. SExtractor is able to perform deblending of close
objects. The number of deblending sub-thresholds   was set to 32 pixels,
and through experimentation, the minimum contrast parameter  was set to 0.0001.
Our catalog contains approximately 40,000 objects per field, 30,000 of which have $I>22.5$mag.

\subsection{Photometry}

 For each object, we measure the color  in a
$2\times$FWHM diameter aperture, where we took seeing variations on different bands  into account:
the color in two bands, e.g.  $(U-B)=m_B(2\times$FWHM$_{B}) - m_U (2\times$FWHM$_U)$, and similarly for other colors.
Although this procedure  is strictly valid only
 for star-like objects, it has been shown to be a good approximation for faint and unresolved galaxies \citep{smail95}.
Indeed, from {\it Hubble Space Telescope}  studies, the half-light radius of LBGs is, on average, 0.4\arcsec\ 
\citep{lowenthal97}, much  less than our seeing.  
For sources that  were not detected in one band (i.e. flux$<1\sigma$), 
the magnitude in that band is set to the $1\sigma$ flux limit in
a $2\times$FWHM diameter aperture and no color term is computed (Equation~1). 
	
In the remainder of this paper, we use aperture magnitudes. To convert
those to total magnitudes, we estimate the total magnitude correction for star-like objects
to be $m_{I}({\rm tot})=m_I(2\times$FWHM$)-0.35$ in the I-band, calibrated by
    adding simulated stars with known total flux into our images and
    measuring the recovered flux in the chosen aperture.

 In addition, each object in our catalogs was corrected for Galactic extinction by adopting $E(B-V)$ values 
taken from the maps of \citet{schlegel98} assuming an  $R_V=3.1$ extinction curve.

\subsection{Completeness}
\label{section:completeness}

In order to estimate our completeness, we added to our images fake stellar objects  with MOFFAT profiles
that matched the image point spread function (PSF). Fluxes  were measured  with the same aperture.
We find that we are 50\% complete up to $m_I\simeq 24.35$mag.
 The exact values for each field are shown in Table~\ref{table:depth}.
Using the transformation
$I_{\rm AB}=m_I+0.47$, this corresponds to $I_{\rm AB}\simeq 24.8$mag ($\mathcal R_{\rm AB}$$\sim25$mag, Steidel et al. 1993)
and to 0.67$L^*$,   where $m_R^*\simeq24.5$ for galaxies at  $z\simeq 3$ \citep{steidel99}.
  Our 90\% completeness level is $I_{\rm AB}\simeq 24.4$mag.
 Thus,   we reached a depth sufficient
to ensure that we sample well  $L^*$ galaxies at $z=3$.

\subsection{Selecting Lyman break galaxy candidates}
\label{section:selection}

From our catalog of $\sim 40,000$ objects, we  rejected   objects
close to the field  edges and objects with a FWHM\_IMAGE less than 0.85\% the FWHM for stellar objects,
presumably cosmic rays or bad pixels. We selected a subsample (1/3) using the following color cuts 
 \begin{eqnarray}
(U-B)_{\rm AB}&>&0.3,  (B-I)_{\rm AB}<3.7, \nonumber \\
  (U-B)_{\rm AB}&>&0.3+0.55\cdot[(B-I)_{\rm AB}-2], \label{eq:colorcut}
 \end{eqnarray} 
and    with   $22.9<I_{\rm AB}\leq 24.8$mag. 
This removed most  $z\leq1$ objects and reduced the amount of computing for the photometric
redshifts, which is explained in the next section.
$\sim 10,000$ objects per field met the criteria shown in Eq.~\ref{eq:colorcut}.

\subsection{Stellar contamination}
\label{section:contamination}

We did not perform any star-galaxy separation since  
beyond $I\simeq23$mag,   number counts are dominated by the faint galaxies at those magnitudes. 
However, we determined empirically the stellar contamination at faint magnitudes the following way.
We first  select point sources, i.e. those with
 FWHM of the source $<1.15\times$FWHM of seeing, up to  $I<23.5$mag
 and that fall in our color selection box (Eq.~\ref{eq:colorcut}).
We then select all the LBG candidates with $22.5<m_I<24.35$mag that have a probability to be at $z=3\pm0.25$ greater
than 50\%, from our photometric redshifts (see below).
  This subset contains some of the point sources between $m_I=22.5$mag  and $m_I=23.5$mag, namely
those with colors consistent with being at $z\simeq 3$. 
Finally, we compare the number counts of the LBG candidates with that of the
point sources that have $z\sim3$ colors and extrapolate the counts at  $m_I>23.5$ to $m_I=24.35$mag,
assuming the counts to be constant at $m_I>23.5$mag (a conservative assumption 
given that the star counts almost certainly  decreases from $m_I=22.5$mag to $m_I=23.5$mag).
We found the stellar contamination to be $<7$\%, $<15$\% and $<13$\% for the APM 08279+5255, the PC 1233+4752, and
the J0124+0044 field, respectively. 
 
\section{Sample selection \& properties}
\label{section:sample}

Following our rough color-cuts, we used photometric redshift techniques to select LBG candidates in 
 narrow redshift slices.   

\subsection{Photometric Redshifts}
\label{section:photoz}

There are two approaches to photometric redshift estimations: the   empirical training set method \citep[e.g.][]{koo85,connolly95}
and the spectral energy distribution (SED) fitting \citep{lanzetta96,sawicki97,budavari00,fontana00,csabai03}.
The former is an empirical relationship between colors and redshifts
 determined using a multi-parametric fit. The latter is based 
 on a set of SED templates (empirical or theoretical).
The two methods are comparable in their performance at $z\leq1$; however, the training
set method is not always feasible, especially at high-redshifts \citep[see discussion in][]{benitez00}.

On the other hand, the  SED  fitting method works best when there is a strong feature in the SED, such as
the  4000\AA\ break, or the 912\AA\ Lyman break. Thus,
SED fitting methods were rapidly developed for the Hubble deep field (HDF)
\citep[e.g.][]{budavari00,fernandez-soto01} with an accuracy of typically $\Delta z\sim 0.06(1+z)$. 
 In our case, we used the code Hyperz from \citet{bolzonella00}, which
 includes intergalactic  absorption due to the \Ly\ forest and internal extinction $A_V$.
 We updated the intergalactic absorption prescription following \citet{massarotti01}
 and we used the extinction curve of  \citet{calzetti00} with $A_V$ varying from 0 to 1.2.

 We used the template set made of the four empirical SEDs of \citet{coleman80}, extended in the UV ($\lambda<1400$\AA)
by \citet{bolzonella00} using the synthetic models of \citet{bruzual93} with parameters (SFR and age)
that matched the spectra at $z=0$. Note that \citet{fernandez-soto01}  extended the CWW templates
using the power laws of \citet{kinney93}, and \citet{budavari00} used the extensions of \citet{kinney96}.
The SED templates are  convolved with the MOSAIC filter response curves (including the CCD response),
and $z_{\rm phot}$ is found from the maximum of the likelihood distribution ${\mathcal L}(z)$
derived from the $\chi^2$ distribution.

\subsection{Simple tests on the HDF}

 We performed several tests of 
 the technique on the sample of 150 spectroscopically confirmed  galaxies  at redshifts  $z\leq6$ in the Hubble deep field north (HDF-N)
   \citep{cohen00,fernandez-soto01}.
We experimented with 5 template sets that included the 4 CWW templates and various starburst templates from Starburst99
\citep{starburst99}. Of the 18 galaxies with  $2.75<z_{\rm phot}<4.5$,  two (11\%) are outliers with  $z_{\rm spec}\simeq 1$.
From that sample, we found that (i) the four CWW templates gave the lowest scatter $\Delta_z/(1+z_{\rm spec})=0.053$,
 both measured using the bi-weight estimator of \citet{beers90}, and (ii)  $\Delta_z/(1+z_{\rm spec})$ is not improved using
near-IR photometry. Thus, near-IR  observations are not required for photometric redshifts at $z\simeq 3$.

\subsection{Using priors}

SED fitting methods give the most likely redshift given the observed set of colors. However,
 information  such as size, or flux,  can be included in photometric redshift techniques using  Bayesian probabilities
 \citep[following][]{benitez00}. We coupled the SED fitting scheme
 with the prior likelihood distribution  for a galaxy of magnitude $m_I$ parametrized by \citet{benitez00}
 in the following way:   the product $prior \times likelihood$ is decomposed over the SED types $T$:
\begin{eqnarray}
P(z)&=&\sum_T p_T(z|m_I)\cdot {\mathcal L}_T(z), \label{eq:prioreq}
\end{eqnarray}
where   $ p_T(z|m_I)$ is  the prior probability given the galaxy magnitude $m_I$, and
${\mathcal L}_T(z)$ is the probability of observing the  galaxy colors if the galaxy is 
at redshift $z$ and has a type $T$. The photometric redshift $z_{\rm phot}$ is taken from
 the maximum of the $P(z)$ distribution,
and the errors, $\sigma_z$, are computed from the FWHM of $P(z)$ divided by $2.35$.
 Redshifts with large $\sigma_z$ may be unreliable.
A good estimator of reliability is the following \citep{benitez00}:
\begin{equation}
P_{\Delta z}\equiv P(|z-z_{\rm phot}|<0.2\times(1+z_{\rm phot}))\label{eq:outlier},
\end{equation}
 which  estimates the `goodness' of   a photometric redshift $z_{\rm phot}$ using Eq.~\ref{eq:prioreq},
 and also has the useful feature to pick likely outliers \citep{benitez00}.
The factor 0.2 is arbitrary, but since the rms of photometric redshifts $\sigma_z$ is $\sim 0.05(1+z)$, this factor
corresponds to approximately $4\times \sigma_z$.
At $z<6$, the overall rms of $\Delta_z$ is 0.11, and $\Delta_z/(1+z_{\rm spec})=0.06$, similar to $0.059$ found by \citet{benitez00}.

\subsection{Photometric redshift distributions}

\begin{figure}
\plotone{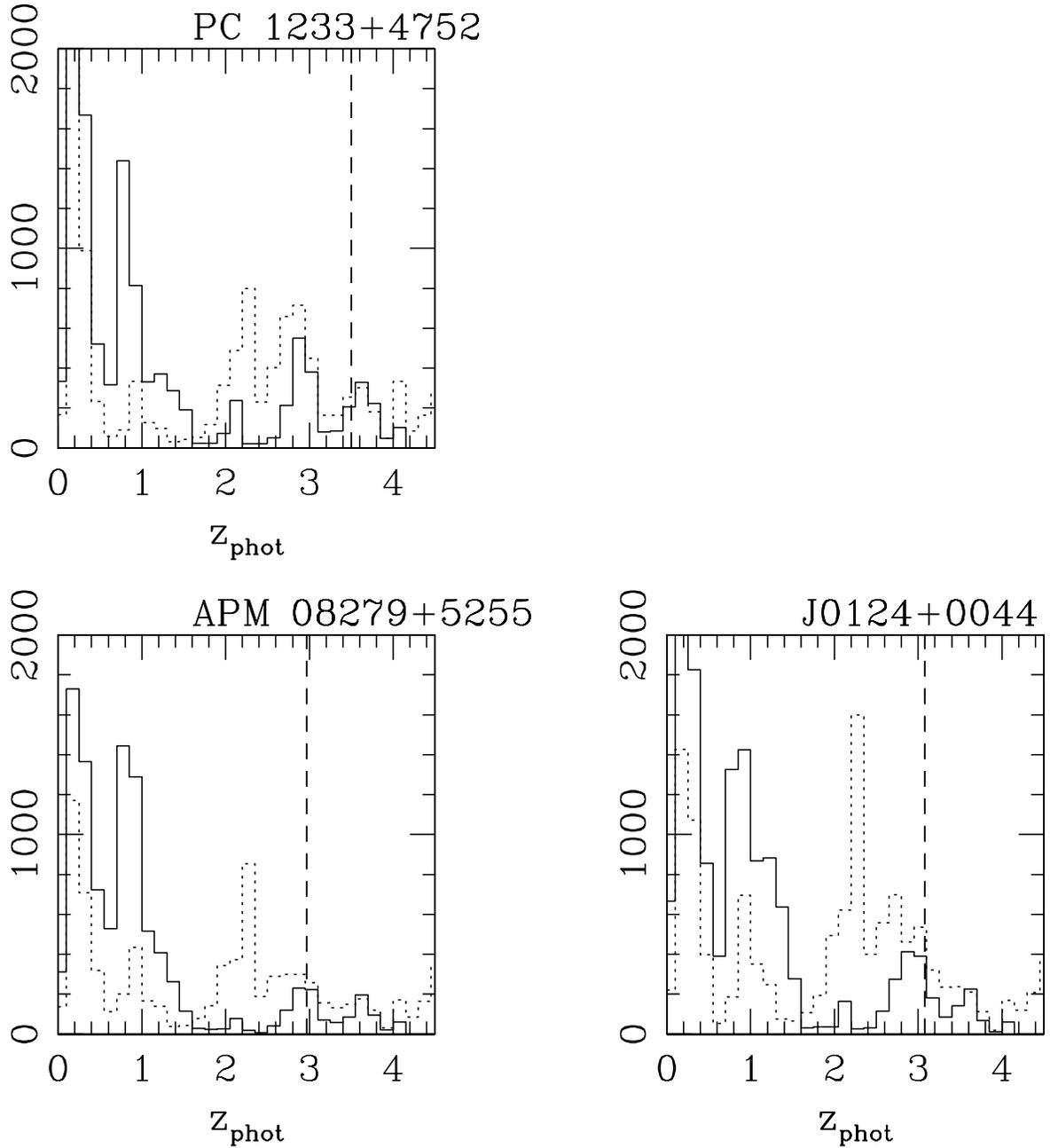}
\caption{Redshift distribution for each of our fields. 
The dotted histogram shows the photometric redshift distribution using no priors and the template set A.
The continuous histogram shows the photometric redshift distribution using the priors.
 Using the priors has the effect
of eliminating the large number of galaxies that have been assigned  $z_{\rm phot}\simeq 2$ wrongly,
but does not affect the distribution at $z\sim 3$ significantly. 
The vertical dashed line shows the redshift of the DLA  $z_{\rm DLA}$. This plot shows the effect
of the priors and that our selection peaks at a redshift close to that of the DLA.
\label{photoz:fig:zdistr}}
\end{figure}

Fig.~\ref{photoz:fig:zdistr} shows the redshift distribution   of the three fields.
The dotted histogram shows the photometric redshift distribution using the CWW templates with no priors.
The continuous histogram shows the photometric redshift distribution using the priors.
Fig.~\ref{photoz:fig:zdistr} shows that using the priors has the effect of removing galaxies with $z_{\rm phot}\simeq 2$ that
are likely at lower redshifts. As expected, the  distribution of galaxies at $z\sim3$ is not affected much.
This is due to the fact that this method is sensitive to the shape of the SED, which has a strong break
between the U and B filters at that redshift.

We used  the photometric redshift $z_{\rm phot}$   of our galaxies to determine their absolute magnitude $M_{I,\rm rest}$.  
The $K$-correction, which for galaxies at $z\sim3$  corresponds to the extrapolation
of their intrinsic flux at $\lambda_{\rm rest}\sim 8000$\AA\ from their observed flux at $\lambda_{\rm rest}\sim 2000$\AA,
was  computed  using a weighted sum on each SED.
 Each template was weighted by the prior probability $p_T(z|m_I)\cdot {\mathcal L}_T(z)$
since the  best-fitted SED was   a combination of the spectral types $T$ (see Eq.~\ref{eq:prioreq}).
At redshift $z\sim3$, the $K$-correction is, however, small: it is typically $\sim 0.2$ for blue SEDs, such as the Irr template
of \citet{coleman80}.

\begin{figure}
\plotone{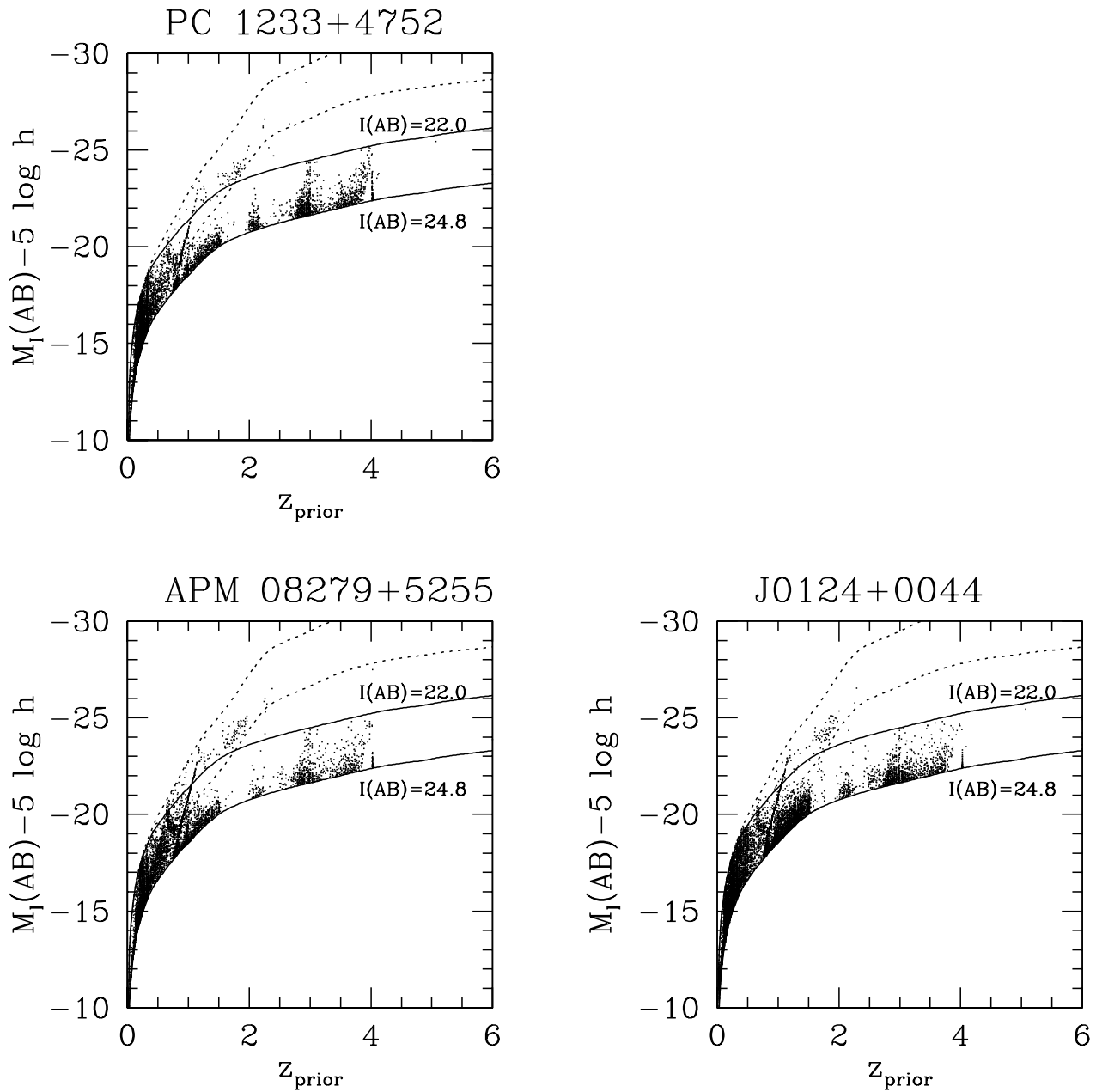}
\caption{Absolute magnitude vs. photometric redshift for each field. Each dot represents one galaxy in our fields.
The rest-frame absolute magnitude $M_I$ was computed using the distance modulus and
 the $K$-correction (see text for details).
The continuous lines show our magnitude selection $22<I_{\rm AB}<24.8$mag and are
computed using an Irr SED.   At $z\sim 3$, galaxies are  
between the two continuous lines and near our completeness limit as expected, which gives us more confidence 
in the photometric redshifts. 
The dotted lines show   our magnitude selection  for an E/S0 SED.
Clearly, points that are outside the range allowed by the continuous lines are
 best-fitted by the E/S0 type, which has a strong break at 4000\AA\  and thus a large $K$-correction.
This plot shows that the photometric redshift technique is self-consistent. 
\label{photoz:fig:plotzM}}
\end{figure}

Fig.~\ref{photoz:fig:plotzM} shows the absolute magnitude $M_I$ as a function of $z$.
Each dot represent one galaxy in our fields. The two continuous lines show our magnitude cuts and were
computed using an Irr SED. The   galaxies at redshift $z\sim 3$ are, as
expected, between the two lines and near our completeness limit,
which provides  a consistency check of the photometric redshift technique.

Almost all points that are outside the range allowed by the continuous lines in Fig.~\ref{photoz:fig:plotzM} are between the dotted lines,
 which represent the magnitude range for an E/S0 template, and thus
are best fitted by the E/S0 type. This type has a strong break at 4000\AA\  
that creates  a large $K$-correction of four magnitudes and makes these objects
 too luminous for their apparent magnitude. The fitted SEDs  are likely to be wrong. 
At that redshift, $z_{\rm phot}\sim 2$, IR photometry is needed to better constrain the SEDs.

\subsection{Selecting reliable galaxies in slices}
\label{section:slices}

From the subsample described in section~\ref{section:selection},
we selected $\sim100$  LBGs that are in a redshift slice centered on the DLA. 
  Specifically, they  were chosen
to have a high probability  of being at the redshift of the DLA, $z_{\rm abs}$, i.e.
\begin{equation}
 P(z_{\rm abs}\pm W_z/2)\equiv \Pdla>0.5,\label{eq:slicecut}
\end{equation}
where $W_z$ is the redshift slice width.
 We also defined two additional redshift slices shifted by $+0.15$ or $-0.15$ from $z_{\rm abs}$,
$\Ppdla$ and $\Pmdla$, respectively.
  We choose a redshift width of $W_z=0.15$ because, as discussed in \citet{bouche03}, it produces
 the largest sample in the smallest redshift slice, given the rms of the photometric redshifts.
 At the end of \S~\ref{section:discussion}, we show that a different choice of $W_z$ does not change the results.
 
\begin{figure}
\plotone{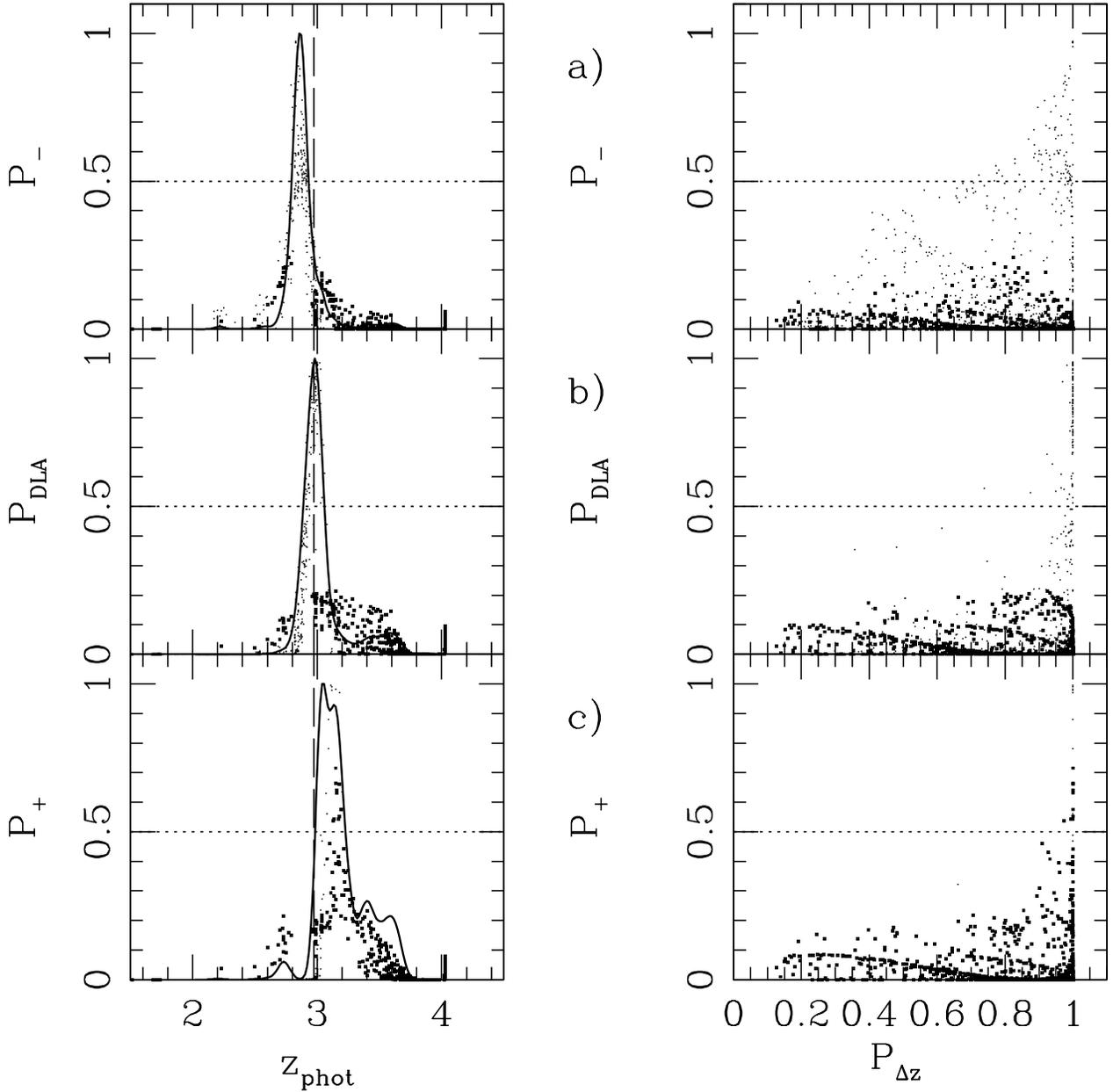}
\caption{Redshift slices centered (a) in front of, (b) on, and (c) behind the   DLA 
for the APM 08279+5255 field. The value 
$z_{\rm abs}$=2.974 is indicated by the vertical dashed line.
Each dot represent a galaxy that was detected in the four UBV\& I bands.
The filled squares indicate objects that are not detected in the U band.
The left column shows the  probability distribution as a function of photometric redshift.
The continuous line shows the   smoothed distribution (arbitrarily scaled to the peak).
The right column shows the  probability to be in that particular slice as a function of the `goodness'
of the photometric redshift $P_{\Delta z}$ defined in Eq.~\ref{eq:outlier}. 
The dotted line shows the minimum threshold (50\%) used in selecting LBG candidates in each of the slices.}
\label{photoz:fig:pdla_z}
\end{figure}

More importantly, this criterion (Eq.~\ref{eq:slicecut}) corresponds to high quality photometric redshifts as
illustrated in Fig.~\ref{photoz:fig:pdla_z} for the APM 08279+5255 field.
The left panels of Fig.~\ref{photoz:fig:pdla_z} show the  probability distribution, and
the right panels show the probability distribution as a function of $P_{\Delta z}$ defined in  Eq.~\ref{eq:outlier}, both
 for the three different redshift slices.
Fig.~\ref{photoz:fig:pdla_z}(a), (b), and (c)  show $\Pmdla$, $\Pdla$ and $\Ppdla$, respectively.
The dots represent galaxies detected in all four bands, UBV\& I.
The filled squares indicate objects that are not detected in the U band.  
Smoothing the distributions using a Gaussian kernel (scaled to the peak)  
 produced the continuous lines in Fig.~\ref{photoz:fig:pdla_z}.
The dotted line shows the minimum threshold ($>0.5$)  used in selecting LBG candidates in each of the slices.
From the right panels, galaxies that have a high probability of being in a redshift slice
 also have reliable photometric redshifts, indicated by the fact that  $P_{\Delta z}\geq 0.9$.
For each redshift slice, the number of galaxies that met the threshold  
 is shown  in Table~\ref{photoz:table:data}.

\begin{figure}
\plotone{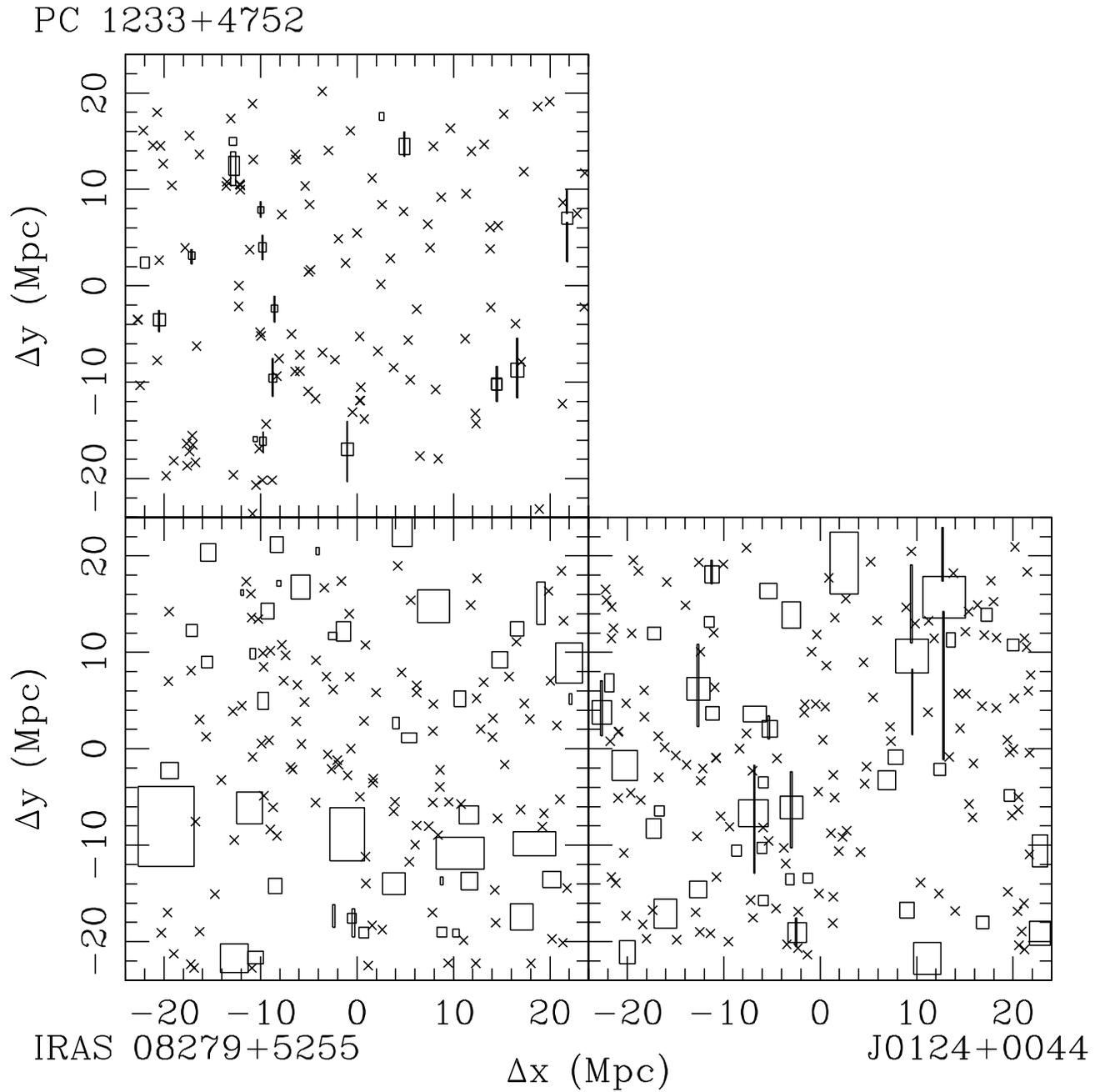}
\caption{For our three fields, the $x$ $y$ position of our LBG candidates relative to the
QSO location. North is left, East is down.} 
\label{fig:xyposition} 
\end{figure}

Fig.~\ref{fig:xyposition} shows the $x-y$ positions of our LBG candidates that met the criterion Eq.~\ref{eq:slicecut}.
The square regions show the masks used to cover bright stars and defects such as streaks.


\section{Clustering analysis}
\label{section:cluster}

 In the current hierarchical theory of galaxy formation \citep[e.g. see][and reference therein]{longair98},
 small quantum fluctuations that were stretched out
to cosmological scales by inflation grew (mainly linearly) during the radiation-dominated era, till
the present. The initial power spectrum ($P(k)\propto k^n$), which characterizes these fluctuations in Fourier space,
is nearly scale invariant (i.e. $n\simeq 1$) on all scales. Initially, all scales grew at the same rate.
Small scales entered the horizon before 
the universe became matter-dominated. During that time their growth   was suppressed.
The resulting power spectrum $P_E(k)$ has $n\simeq 1(-3)$ on large (small) scales.
  These dark-matter fluctuations  formed deep  gravitational potentials in which galaxies and
galaxy clusters  formed. When the density contrast reached $\delta \rho/\rho\sim 1$, the fluctuations grew non-linearly
until $\Delta \rho/\rho\sim 200$. 

Since only gravity is driving this build-up of matter,
 massive galaxies are more likely to be found in high-density regions, whereas low-mass galaxies are more uniformly
distributed. This produces an enhancement of the clustering of massive galaxies.
 Therefore, the  clustering properties of galaxies probe their dark-matter mass distribution. 
The auto-correlation $\xi(r)$ is a natural tool to study  clustering   in this context, since
$\xi_{\rm DM}$ is the Fourier transform of the evolved power spectrum $P_E(k)$ \citep[e.g.][]{peacock99}. 
The galaxy correlation $\xi_{gg}$ is related to the dark-matter auto-correlation $\xi_{\rm DM}$ via
the bias $b$. At a given redshift,
\begin{equation}
\xi_{gg}(r)=b^2(M)\xi_{\rm DM}(r). \label{eq:biasCDM}
\end{equation}
This bias can be computed in the   Press-Schechter formalism extended by \citet{mo02} and reference therein.

Similarly to the galaxy auto-correlation $\xi_{gg}$, one can define the cross-correlation $\xi_{dg}$
between DLAs and LBGs 	from the conditional probability of finding a galaxy in a volume d$V$ at a
distance $ r=|\mathbf r_2-\mathbf r_1|$, given that there is a DLA at $\mathbf r_1$:
\begin{equation}
 P(LBG|DLA)	= n_u (1+\xi_{dg}(r)) \mathrm d V_2, \label{eq:cross}
\end{equation}
where $n_u$ is the unconditional background galaxy density.
Thus, the   number of neighbor galaxies  in a cell of volume $\Delta V$  is given by
$  N_p= \overline N (1+\overline \xi_{dg}(r)),$
where $\overline N=n_g\Delta V$ and $\overline \xi$ is the cross-correlation averaged over the volume $\Delta V$.
This estimator of $\xi$ requires an estimate of the unconditional background galaxy density $n_g$.
 There are  two ways to quantify $n_g$ in case of the cross-correlation: 
 one way is to use galaxies spatially far from the DLA as in \citet{bouche03}; the other way
 is to use  the entire galaxy catalog. Here, we used the latter because of the simplicity
 of this method when randomizing the line of sight (see section~\ref{section:res:errors}).
Naturally, large fields will yield a better estimate of $n_g$.

We can extend the spatial cross-correlation $\xi$ to angular correlation function $w$ since
    the former is directly related to spatial correlation functions $\xi$.
 Namely, if the selection function $\phi(z)$ $\simeq 1/W_z$ within $[-W_z/2,W_z/2]$ and zero otherwise, 
 then 
 \begin{equation}
 w(r)=\frac{2}{W_z}\int_0^{W_z/2} \xi\left(\sqrt{z^2+r^2}\right) \mathrm d z,\label{eq:omegaZ}
 \end{equation}
or $w( r_\theta )\simeq  \frac{A}{W_z} \left ( \frac{r_\theta}{r_o} \right )^{\beta}$
where $\beta=1-\gamma$ and $A$ is a constant that can be computed analytically \citep[e.g.][]{adelberger03,eisenstein03}.

We used the  following  estimator of the cross-correlation $w_{dg}(r)$,
\begin{eqnarray}
1+\overline w_{dg}(r)&=& \left < \frac{N_r}{N_g}\frac{N_{\rm obs}(r)}{N_{\rm rand}(r)} \right >, \label{eq:estimator}
\end{eqnarray}
where  $N_{\rm obs}(r)$ is  the observed number of galaxies   between $r-dr/2$ and $r+dr/2$,
 $N_{\rm rand}(r)$ is the number of randomly distributed galaxies,
 $N_g$  is the total number of galaxies,   $N_r$ is the total number of random galaxies, and
$<>$ denotes the   average   over the number of DLAs ($N_{\rm DLA}$).
The random catalog $N_r$ is  much larger than the galaxy catalog  
so that the variance of  $N_{\rm rand}(r)$ is negligible.

The errors to  $w$  in Eq.~\ref{eq:estimator} are given by \citep[e.g.][]{landy93}
\begin{eqnarray}
\sigma_{w}&\simeq &\frac{\sigma_{<N_{\rm obs}>}}{<N_{\rm obs}>} [1+\overline w_{dg}]  ,  \label{eq:xi:variance}
\end{eqnarray}
where $<N_{\rm obs}>$ is the number of neighbors in the annuli  averaged over $N_{\rm DLA}$. 
Using solely the Poisson variance of the mean $<N_{\rm obs}>$, i.e.
 $\sigma_{<N_{\rm obs}>}=\sqrt{{N_{\rm obs}}/{N_{\rm DLA}}}$, yields  
the Poisson errors for  Eq.~\ref{eq:estimator} \citep[e.g.][]{mo92,landy93}: 
\begin{eqnarray}
\sigma_{w}&\simeq &\frac{1+\overline w_{dg}}{\sqrt{N_{\rm DLA}<N_{\rm obs}>}}   . \label{eq:xi:Poisson}
\end{eqnarray}
In general, the cross-correlation and the galaxy auto-correlation will increase the variance
to $<N_{\rm obs}>$ , and   $\sigma_{w}$ will be larger \citep[see][]{eisenstein03}. 

Since $<N_{\rm obs}>$ is proportional to the total number of galaxies, $N_g$,  
 the expected rms of the cross-correlation function is 
\begin{equation}
 \sigma_w \propto \frac{1}{\sqrt {N_{\rm DLA}N_g}}  \label{eq:xi:err}
\end{equation}
as one might have expected.
Thus, the noise in  $\overline w_{dg}$ goes as the inverse of the square
root of the number of DLAs, $N_{\rm DLA}$, and as the inverse of the square root of
the number of galaxies  $N_g$.


 \section{Results}
\label{section:results}

With the clustering formalism layed out in section~\ref{section:cluster}, we can
present our results on the DLA-LBG cross-correlation (\S~\ref{section:res:cross})
and on the computation of the errors (\S~\ref{section:res:errors}).

 \subsection{DLA-LBG cross-correlation from the combined fields}
\label{section:res:cross}

\begin{figure}
\plotone{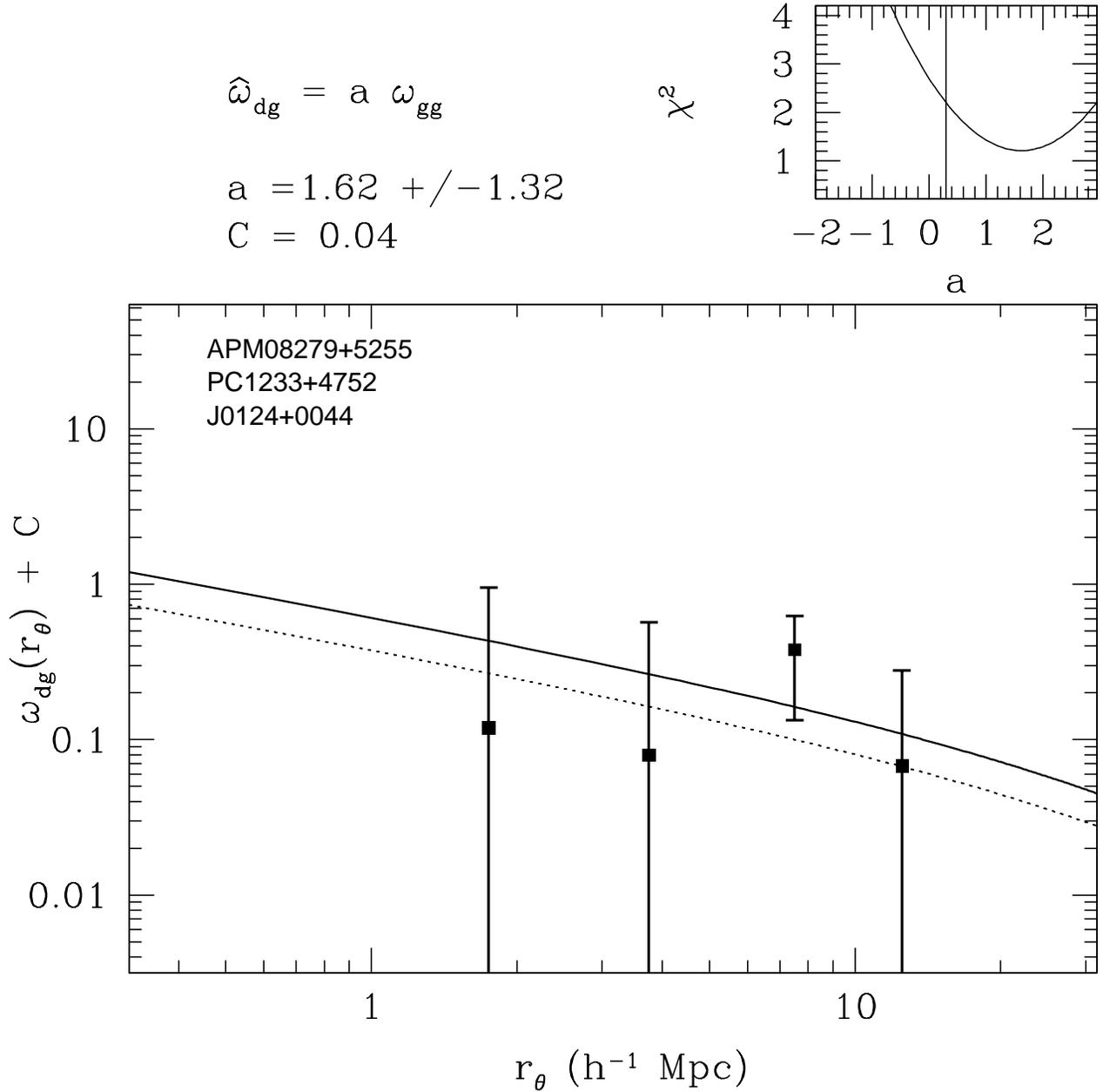}
\caption{The cross-correlation $w_{dg}$ between  DLAs and Lyman break galaxies in
a redshift slice of   width ($W_z=0.15$) that contains the DLAs.
The  filled squares show the  cross-correlation for the combined fields.
The dotted line is the LBG auto-correlation $w_{gg}$ 
\citep[from][using Eq.~\ref{eq:omegaZ} to account for the volume of our redshift slice]{adelberger03}.
The continuous line is a  fit to the amplitude of the cross-correlation using 
$\hat w_{dg}=a\times w_{gg}$, i.e. we assume that both $w_{gg} $ and $w_{dg}$ have the same slope.
The small panel shows the $\chi^2$ distribution as a function of the amplitude $a$ and the $1\sigma$ range.
\label{fig:xcorrelDLA:average:Pdla}}
\end{figure}

Fig.~\ref{fig:xcorrelDLA:average:Pdla} shows the DLA-LBG cross-correlation $w_{dg}$
computed using Eq.~\ref{eq:estimator}. In computing $N_{\rm obs}(r)$ and $N_{\rm rand}$ in Eq.~\ref{eq:estimator},
 we  took into account the masked regions with bright stars shown in Fig.~\ref{fig:xyposition}.
 The dotted line shows the auto-correlation $w_{gg}$
of \citet{adelberger03} and the continuous line shows a fit to the amplitude of $w_{dg}$
using $w_{gg}$ as template. The fitting method is described below.

It is necessary to take into account the different selections of the different fields
in performing  the sum in Eq.~\ref{eq:estimator} indicated  by the brackets, so
we must weight each field accordingly.
We chose to weight each field according to its errors   at each angular scale $r_i$.
Thus, for each field $l$, we compute $\frac{N_{\rm obs}}{N_{\rm rand}}\equiv 1+w_l(r)$
in six annuli. Then,
the combined angular cross-correlation, $w_{dg}$  is computed from the weighted mean of $w_l(r_i)$,
 where the weights were  $\sigma_l(r_i)$ at each angular scale $r_i$.
 
 The error bars shown in Fig.~\ref{fig:xcorrelDLA:average:Pdla}, $\sigma(r_i)$, are from
the full covariance matrix to $w_{dg}$. The latter is found using the standard error propagation formula:
 \begin{equation} 
 {\rm COV}_{i,j}= \sum_{l=1}^{N_{\rm DLA}} \D{f(r_i)}{w_l(r_i)} \D{f(r_j)}{w_l(r_j)} {\rm COV}_l(r_i,r_j)\label{eq:cov:aver}
 \end{equation}
where $N_{\rm DLA}$ is the number of DLAs, $f(r)$ is the weighted mean $w_{dg}(r)$
 and ${\rm COV}_l(r_i,r_j)$ is the covariance matrix of the individual field $l$  given by Eq.~\ref{eq:COV:field} (below).

\subsection{Error computation}
\label{section:res:errors}

Because each DLA is at a slightly different redshift,   each field has a different selection
function, and  it is necessary to take into account these differences 
by weighting  each field accordingly. As mentioned,
we chose to weight each field according to its errors $\sigma_w(r_i)$.

The errors need to be computed carefully.  Several options are available.
 The proper way to compute the errors would be to resample the DLAs (via bootstrap techniques), 
 but this is impractical here given the number of DLA fields at our disposal.
Another way would be to bootstrap the galaxies, which would
 reproduce only the Poissonian errors (Eq.~\ref{eq:xi:Poisson}). 

We used yet another method, which is to perform Monte Carlo simulations in which we randomize the position of the DLA. 
This takes into account the clustering variance due to the galaxy auto-correlation,
 but misses the variance (and co-variance) due to the cross-correlation itself
(the factor $1+w_{dg}$ in Eq.~\ref{eq:xi:variance}). 
However, this term will be small on scales larger than $5h^{-1}$~Mpc because  $w_{dg}<<1$.

Thus, in each DLA field $l$,  we computed  the full covariance matrix ${\rm COV}_l$ from 
 $n_r=200$ randomizations of the DLA position:
\begin{equation}
{\rm COV}_l(r_i,r_j)=\frac{1}{n_r-1}\sum_k^{n_r} [w_k(r_i)-\overline w(r_i)]
[w_k(r_j)-\overline w(r_j)]\label{eq:COV:field}
\end{equation}
where $w_k$ is the $k$th measurement of $w$ and $\overline w$ is the average of the $n_r$
measurements of the cross-correlation.
 The errors $\sigma_l(r_i)$ to  $w_l(r_i)$ for each field $l$ follow.

Our errors  are  consistent with the  errors expected from our analysis
 of cosmological  simulations: in \citet[][in preparation]{bouche03b}, we conclude  that
 with a  data set of this size, we can be sensitive to the cross-correlation   
only on scales 5-10$h^{-1}$~Mpc,  which is where we see a positive cross-correlation.

\subsection{The integral constraint}
\label{section:intconstraint}

Because the unconditional  galaxy density, $n_u$ in Eq.~\ref{eq:cross},  
is estimated from the total observed galaxy density, whereas it
should  always be lower than the observed galaxy density,  all estimates of $\xi$ (or $w$) 
are biased low. 
 This bias $\Delta w$,  often referred to as the `integral constraint',
  can be computed analytically \citep[e.g.][]{landy93,saslaw00}.
  For the angular cross-correlation function, it is :
\begin{equation}
\Delta w=\frac{1}{\Omega}\int {\mathrm d} \Omega \; \hat w_{dg}(r_\theta),
\label{eq:xi:bias}
\end{equation}
where $\hat w_{dg}$ is a model of the cross-correlation and $\Omega$ is the total survey area.
Thus, $\Delta w$ will be smaller with larger fields, and 
we  corrected $w_{dg}$ in Fig.~\ref{fig:xcorrelDLA:average:Pdla}
for its bias $C$ due to the integral constraint
  assuming that $w_{dg}$ is $w_{gg}$ from \citet{adelberger03}.

\section{Discussion}
\label{section:discussion}

We next discuss quantitatively how significant the cross-correlation shown in  Fig.~\ref{fig:xcorrelDLA:average:Pdla}
is. We also show that our measurement of the DLA-LBG cross-correlation (i) is not reproduced by random lines of sight
(at the 95\% level) and  (ii) is not seen in other redshift slices that do not contain the DLAs. 
We compare these results to past and future studies at the end of this section.

\subsection{Is this result consistent with no clustering?}

It is clear from Fig.~\ref{fig:xcorrelDLA:average:Pdla} that we
will not be able to constrain the slope of the cross-correlation, so
 we assumed that the cross-correlation $w_{\rm dg}$ has the same shape 
as the auto-correlation $w_{\rm gg}$. That is, we fitted the cross-correlation
to the model
\begin{equation}
\hat w_{\rm dg}=a\times \hat w_{\rm gg}, \label{eq:omega:model}
\end{equation} where $a$ is the unknown amplitude.

We fitted the amplitude of the cross-correlation $a$ using the covariance ${\rm COV}_{i,j}$
computed in Eq.~\ref{eq:cov:aver}, using the following $\chi^2$ statistic:
\begin{equation}
\chi^2=\sum_{i,j}[w_{\rm dg}(r_{\theta_i})-\hat w_{\rm dg}(r_{\theta_i})]{\rm COV}^{-1}_{i,j} 
		[w_{\rm dg}(r_{\theta_j})-\hat w_{\rm dg}(r_{\theta_j})],
\label{eq:xi:cov}
\end{equation}
where $\hat w_{\rm dg}$ is the model of the angular cross-correlation and ${\rm COV}^{-1}$
is the inverse of the covariance matrix computed using a Single Value Decomposition algorithm.
The $\chi^2(a)$ distribution is shown in the small panel in Fig.~\ref{fig:xcorrelDLA:average:Pdla}.
 In performing the inversion of the covariance
matrix, we rejected   the eigenvalue corresponding to the last radial bin or to
 scales similar to the size of the field  where there is no signal
(see discussion in Bernstein [1994]).

We found that an amplitude of $a>0 $ was favored:
\begin{equation}
a=\crossPdla \pm \crossPdlaerr.\label{xcorrel:results}
\end{equation}

This  measurement of the DLA-LBG cross-correlation is most significant on scales greater
than 5-10$h^{-1}$~Mpc. At this point, we cannot, however, conclude whether   the DLA-LBG cross-correlation is 
stronger or weaker than the LBG-LBG auto-correlation. Taking $r_o\simeq 4h^{-1}$~Mpc and $\beta \simeq 0.6$ for the LBG auto-correlation
\citep{porciani02,adelberger03}, we find that the correlation length of the cross-correlation $w_{dg}$
is $r_o=5\pm 4.5h^{-1}$~Mpc (or $r_o=7\pm 6.8h_{71}^{-1}$~Mpc).

\subsection{Is this result drawn from random lines of sight?}

Given the large rms to the fitted amplitude $a$, could  our result simply be a large fluctuation of the set
of possible values for random lines of sight?
 To test this, we chose 100 lines of sight selected at random,
 excluding the central $5h^{-1}$~Mpc to ensure that the new lines of sight are not 
 correlated with the ones centered on the DLAs.
 We then computed the cross-correlation for these 100 random lines of sight
in the redshift slices centered on the DLAs. As before, we computed the weighted mean
to $w_{dg}$ and used Eq.~\ref{eq:cov:aver}.

\begin{figure}
\plotone{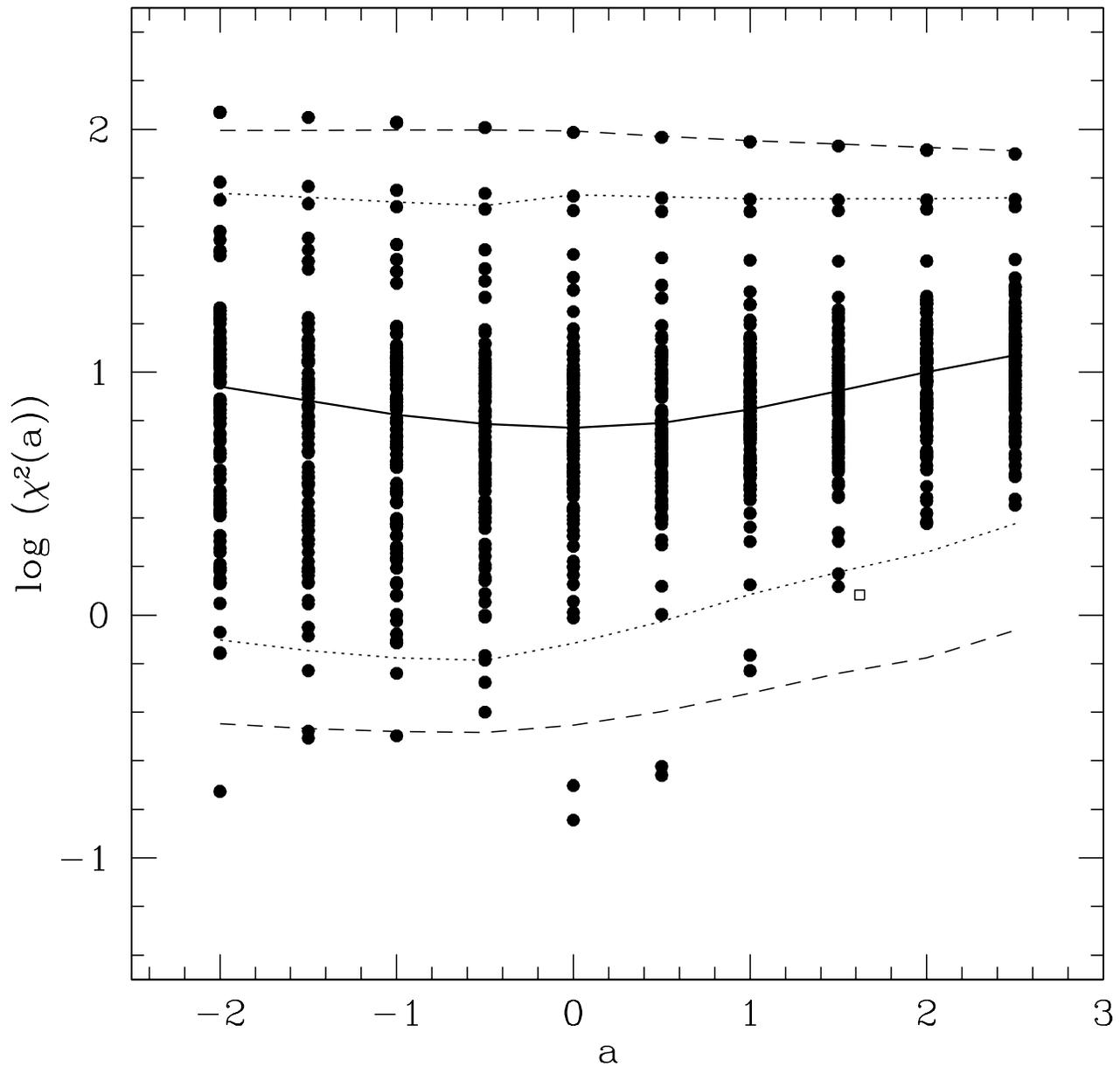}
\caption{The   points show the 
logarithm of $\chi^2(a)$ as a function of the  amplitude $a$ for 100 random lines of sight
(excluding the central 5$h^{-1}$~Mpc) in the redshift slice centered on the DLAs.
 The continuous line shows the median of the  distributions. 
The dotted line and the dashed line are the 95\% and 99\% confidence levels, respectively.
The open square shows the  location of the fitted amplitude shown in Fig.~\ref{fig:xcorrelDLA:average:Pdla}, which 
shows that the signal measured in Fig.~\ref{fig:xcorrelDLA:average:Pdla} is not drawn from   a random
distribution of lines of sight at the $> 95$\% confidence level.
\label{fig:plot_chi}}
\end{figure}

 Fig.~\ref{fig:plot_chi} shows the logarithm of the $\chi^2(a)$ for fixed amplitudes $a$
for the 100 random lines of sight (filled circles).
  The continuous line shows the median of the  distributions. The dotted   and dashed lines
are the 95\% and 99\% levels of the distributions.
  The median, 95\% and 99\% levels are found after a Gaussian kernel smoothing of the distributions
  using the optimum band width \citep{wand95} (The results are not significantly 
  changed using a fixed band width).
The open square shows the location of the result of Fig.~\ref{fig:xcorrelDLA:average:Pdla}.
Since it lies close to the 95\% confidence level,
this shows that the signal measured in Fig.~\ref{fig:xcorrelDLA:average:Pdla} 
is not drawn from  a  random distribution of lines of sight, at the $>95$\% confidence.

\subsection{How about other redshift slices?}

The result of Fig.~\ref{fig:xcorrelDLA:average:Pdla} should be compared with the cross-correlation when there 
is no DLA in the redshift bin. From
our photometric redshift analysis, we selected galaxies in two other redshift slices that did NOT
contain the DLA. We chose the  slices that were in the foreground and in the background
from the DLA, and offset by  $+$ or $-$0.15 in redshift (see Fig.~\ref{photoz:fig:pdla_z}).
In each case, the $\chi^2$ fit does not favor any clustering:
the best amplitude is $a=\crossPm \pm \crossPmerr$ and $a=\crossPp \pm \crossPperr$, respectively.  
A clustering signal in this slice would have cast a strong doubt on our results that do
contain the DLA in Fig.~\ref{fig:xcorrelDLA:average:Pdla}.
In addition, we performed the same check on another slice at redshift $3.6$. The best
amplitude for this slice is $a=\crossPl \pm \crossPlerr$.

We repeated the analysis with $W_z=0.20$ to test whether the observed clustering depends on the choice of 
the slice width. We found
that $a=\crossPddla \pm  \crossPddlaerr$ in this case, so we conclude that the slice width 
does not strongly affect the clustering signal.

\subsection{Comparison with past and future work}

 \citet{wolfe93}  also  found that Ly-emitters are strongly clustered around DLAs.
In contrast, \citet{gawiser01} did not find evidence of clustering and 
the study of \citet{adelberger03}   found a lack of  galaxies near their four DLAs,
within $5.7h^{-1}$~Mpc. Since these two surveys were not sensitive to clustering on  scales larger than $>5h^{-1}$~Mpc,
and ours is not sensitive to $<3-5h^{-1}$~Mpc, our results are not inconsistent with theirs.
The lack of galaxies on small scales could, however, be due to more local environmental effects, such as strong galactic winds from
star forming galaxies.

Although simulations of DLA properties exist (e.g. \citet{katz96,gardner01,nagamine03}), 
no prediction of the DLA-LBG cross-correlation has been published.
In \citet[][in preparation]{bouche03b}, we use  the Tree-Smoothed Particle Hydrodynamical (TreeSPH) cosmological simulations of \citet{katz96a}
to measure the `theoretical' DLA-LBG cross-correlation. These simulations
 contain 128$^3$ dark matter particles and as many gas or star particles.
Each galaxy ($\equiv$   $>64$SPH bound particles) is able to form stars. 
With a  similar number of DLA-LBG pairs and redshift depth ($W_z=111h^{-1}$Mpc),
 we find  $w_{dg}>0$ with the same signal to noise.  Furthermore,
 with a much larger sample of 200 simulated DLAs, 
 we find $w_{dg}\simeq 0.75\pm 0.1$, or $r_o\simeq 3.5h^{-1}$~Mpc.  Thus, DLA halos are less massive than the halos
of LBGs, which are  $10^{12}$\msun\ \citep{porciani02,ouchi03}. 
Given the present sample of three DLAs, our observed constraint on $w_{dg}$ with its   uncertainty
 is   consistent with these simulation results.

\section{Summary and Conclusions}
\label{section:conclusions}

Based  on deep ($\mu_{I,\rm AB}(sky) \simeq 27.6$~mag~arcsec$^{-2}$) wide-field images (0.31$\deg^2$ or 
$\sim65\times 65 h_{71}^{-1}$~Mpc co-moving
 at redshift $z=3$) around three DLAs, we  identify   LBG candidates  brighter than $I_{AB}=\complAB$mag
  using photometric redshift techniques
that included the   $I$ magnitude as a prior estimate in addition to the colors.

From the redshift likelihood distributions, we selected LBG galaxies 
within a redshift slice of width $W_z=0.15(\simeq\sigma_z)$ centered on the redshift of the  DLAs $z_{\rm abs}$. 
Within that slice, we cross-correlated the LBGs with the position of the DLAs
and found that  
\begin{itemize}
\item the amplitude of the DLA-LBG cross-correlation $w_{dg}$ relative to the auto-correlation $w_{gg}$
was $w_{dg}/w_{gg}\equiv a=\crossPdla \pm \crossPdlaerr$, corresponding to  $r_o=5\pm 4.5h^{-1}$~Mpc (co-moving),
\item the amplitude of the DLA-LBG cross-correlation is  $a>0$, which is  significant at the $>95$\% confidence level
based on Monte Carlo simulations,
\item the clustering signal was not present in three redshift slices that did not contain the DLAs.
\end{itemize}

Given the uncertainty of our results, we
 cannot put constraints on the halo masses of DLAs and discriminate between the large disk hypothesis \citep[e.g][]{wolfe86},
  and small sub-L$^*$ hypothesis \citep[e.g][]{maller00,haehnelt00,moller02}.
 Our observation of the clustering on large scales ($>4h^{-1}$~Mpc) is not inconsistent with previous clustering studies 
\citep{gawiser01,adelberger03} since these were limited to small scales.
In order to be able to directly compare these studies with our present results on scales $<4h^{-1}$~Mpc,
a larger sample of DLAs  and multi-object spectroscopy of our LBG candidates are needed.
This will enable to test whether the cross-correlation is stronger or weaker than the auto-correlation.

\acknowledgments
N.B. acknowledges  a post-doc fellowship from the European Community Research and Training Network 
 ``The Physics of the Intergalactic Medium''. J. D. L.
 acknowledges support from NSF grant AST-0206016. 
We  thank the anonymous referee for a careful reading of the manuscript that
improved the quality of the paper. We also thank H. Mo, N. Katz and B. M\'enard for helpful discussions,
and  J. Fynbo for reading a earlier draft.

\clearpage

\begin{deluxetable}{lrrr}
\tablewidth{0pt}
\tablecaption{Properties of the  DLAs and the  QSOs.\label{table:fields} }
\tablehead{ 
 \colhead{}     & \colhead{APM 08279+525}        		& \colhead{PC 1233+4752}         	   & \colhead{J0124+0044}	 
} 
\startdata 
R.A. (J2000) \dotfill   &  $08^h31^m41.6^s$      		&  $12^h35^m31.1^s$    			   &   $01^h24^m03.8^s$ 
          \\ 
Dec. (J2000)  \dotfill   &  $52\arcdeg~45\arcmin~17\arcsec$	& $47\arcdeg~36\arcmin~06\arcsec$          &  $00\arcdeg~44\arcmin~33\arcsec$  
                \\ 
$l$ \dotfill		&  165$\arcdeg~45\arcmin~17\arcsec$	& $130\arcdeg~32\arcmin~12\arcsec$	   &	   $139\arcdeg~58\arcmin~19\arcsec$ \\
$b$ \dotfill		&   36$\arcdeg~14\arcmin~25\arcsec$	& $69\arcdeg~17\arcmin~27\arcsec$	   &	   $-61\arcdeg~02\arcmin~43\arcsec$ \\
$A_U$~\tablenotemark{a} (mag) \dotfill  & 0.20          	& 0.08         				   & 0.13 \\
$E_{B-V}$	\dotfill &  0.04				&	0.02				   & 0.03 \\   \hline
 & \multicolumn{3}{c}{QSO Properties} \\ 
 \hline
 QSO	\dotfill	&  BAL 					&					& 		\\
 $z_{QSO}$  \dotfill    & 3.81         				& 4.447         			& 3.840         \\ 
$m_R$    \dotfill	& 15.2\tablenotemark{1}       		&  20.63\tablenotemark{1}       	&     17.9\tablenotemark{1}         \\  
Radio   \dotfill	& $S_{20\rm cm}=1.3$mJy~\tablenotemark{2}&$S_{6\rm cm}<77\mu$Jy \tablenotemark{3}&     $S_{20\rm cm}=0.11$mJy~\tablenotemark{4}  \\
\hline 
 & \multicolumn{3}{c}{ DLA properties}   \\ 
 \hline 
 $z_{\rm abs}$    				\dotfill & 2.974        		& 3.499\tablenotemark{5}        &  3.077\tablenotemark{7}            \\ 
\HI \hfill $W_r$ (\AA)				\dotfill & $>4.8$\tablenotemark{6} 	& 4.22\tablenotemark{5} 	&  ...\\

\multicolumn{1}{r}{$\log N_{\HI}$ (cm$^{-2}$) } \dotfill &  $<20.3$\tablenotemark{6}   	&  20.9\tablenotemark{5}        &  20.1\tablenotemark{7}           \\		
$ [\Fe/\HH] $       				\dotfill &  -2.31\tablenotemark{6}      &  ...     		        &    ...	\\
\enddata 
\tablenotetext{a}{Galactic extinction from \citet{schlegel98}, averaged over the field.}
\tablerefs{
(1)~\citet{veron01};
(2)~\citet{mcmahon02}
(3)~\citet{schneider91}
(4)~\citet{carilli01}
(5)~\citet{white93}; 
(6)~\citet{petitjean00}; 
(7)~P\'eroux, C., 2003, private communication.
}
\end{deluxetable}

\begin{deluxetable}{lrrrrr}
\tablewidth{0pt}
\tablecaption{Summary of the observations. \label{table:observations}}
\tablehead{ 
	\colhead{} &	
	\colhead{U band} &
	\colhead{B band} &
	\colhead{V band} &
	\colhead{I band} &
	\colhead{} \\ \hline
\colhead{Field}      & \multicolumn{4}{c}{ Total Exposure Time } & \colhead{UT Date of Obs.}
}
\startdata
APM 08279+5255  &  3.75hr & 35min   & 50min  & 2.08hr &  Feb. 7, 8, 2000 \\
PC 1233+4752    &  3.50hr & 40min   & 50min  & 1.92hr &  Feb. 7, 8, 2000 \\
J0124+0044      &  3.72hr & 47min   & 52min   & 2.08hr & Sept. 23--26, 2001
\enddata
\end{deluxetable}


\begin{deluxetable}{lr|rrr}
\tablecaption{Photometric solution. 
\label{table:zeropoint}}
\tablewidth{0pt}
\tablehead{ 
	\colhead{}   &	
	\colhead{Run I \tablenotemark{a}	} 	&
	\multicolumn{3}{c}{Run II	} \\
	    \colhead{Filter	} 	&
	    \colhead{ZP 	}     	&
	    \colhead{ZP 	}     	&
	    \colhead{$\alpha$	}	&
	    \colhead{$\beta$	}
}
\startdata
 U  &  23.26 (0.08) 	& 23.52 (0.02)   & -0.421 (0.02)   &   0.018 (0.007)  \\
 B  &  25.085 (0.03) 	& 25.26 (0.02)   & -0.180 (0.01)   &   0.095 (0.008)  \\
 V  &  25.07 (0.02)	& 25.25 (0.02)   & -0.072 (0.01)   &  -0.025 (0.009)  \\
 I  &  24.58 (0.05) 	& 24.76 (0.02)   &  0.020 (0.01)   &   0.003 (0.011) 
\enddata
\tablenotetext{a}{For run I, we assumed the airmass coefficient $\alpha$ and the
color term $\beta$ to be the same as for run II.}
\end{deluxetable}


\begin{deluxetable}{lcccccccc}
\tablecaption{Depth of the observations \label{table:depth} }
\tablewidth{0pt}
\tablehead{
	\colhead{} &
	\colhead{} &
	\colhead{Exp. /Frames} &
	\colhead{Airmass $X$\tablenotemark{a}} &
	\colhead{FWHM} &
	\colhead{SB$_{lim}(1\sigma$) \tablenotemark{b}	} &
	\colhead{SB$_{lim}(5\sigma$) \tablenotemark{b}	} &
	\colhead{m$_{lim}(3\sigma$)  \tablenotemark{c}	} &
	\colhead{Completeness 50\% \tablenotemark{c}} \\
	\colhead{Fields}  &
	\colhead{Filter} &
	\colhead{(sec./\#)} &
	\colhead{(min-max)} &
	\colhead{(arcsec)} &
	\colhead{(mag/m$_{AB}$)} &
	\colhead{(mag/m$_{AB}$)} &
	\colhead{(mag/m$_{AB}$)} &
	\colhead{(mag/m$_{AB}$)}
}
\startdata

APM 08279+5255  & U & 13500/15 	& 1.07-1.17 & 1.1 & 27.78/28.49 & 26.03/26.74 & 25.85/26.56  & \\
		& B &  2100/7   & 1.14-1.27 & 1.1 & 28.44/28.36 & 26.69/26.61 & 26.49/26.42  & \\
		& V &  3000/10  & 1.21-1.37 & 1.2 & 28.21/28.23 & 26.46/26.48 & 26.17/26.19  &  \\
		& I &  7590/20  & 1.11-1.62 & 1.1 & 27.15/27.61 & 25.40/25.86 & 25.21/25.66  & 24.40/24.87 \\ \hline
	
PC 1233+4752    & U & 12600/14  & 1.04-1.16 & 1.05& 27.82/28.52 & 26.07/26.78 & 25.94/26.65  & \\
		& B &  2400/8   & 1.05-1.07 & 1.0 & 28.59/28.51 & 26.84/26.76 & 26.75/26.68  & \\
		& V &  7590/10  & 1.07-1.11 & 0.9 & 28.18/28.20 & 26.43/26.45 & 26.45/26.47  & \\
		& I &  6900/15  & 1.07-1.45 & 1.1 & 27.19/27.64 & 25.44/25.90 & 25.30/25.76  & 24.48/24.95\\ \hline
	
J0124+0044      & U & 13400/16  & 1.19-1.51 & 1.5 & 27.88/28.59 & 26.13/26.84 & 25.60/26.31 & \\
		& B &  2800/7   & 1.24-1.42 & 1.5 & 28.69/28.61 & 26.94/26.87 & 26.44/26.37 & \\
		& V &  3100/7   & 1.46-1.90 & 1.4 & 28.33/28.35 & 26.58/26.60 & 26.14/26.16 & \\
		& I &  7500/20  & 1.21-1.80 & 1.1 & 27.67/28.13 & 25.93/26.38 & 25.75/26.21 &  24.48/24.95 \\ \hline
\enddata
\tablenotetext{a}{The airmass is $\frac{1}{\cos \zeta}$ where $\zeta$ is
the zenith angle of the telescope.}
\tablenotetext{b}{Limiting surface brightness in magnitudes per square arcsecond.}
\tablenotetext{c}{Measured inside a $2\times$ FWHM diameter aperture.}
\end{deluxetable}

\begin{deluxetable}{lcccc}
\tablewidth{0pt}
\tablecaption{Number of galaxies in the different redshift slices. \label{photoz:table:data} }
\tablehead{
\colhead{Field} &\colhead{$\Pmdla$}& \colhead{$\Pdla$} &\colhead{$\Ppdla$} 
}
\startdata
APM 08279+5255 &	89	& 84	&  17     \\
PC 1233+4752	&	20	& 70	&  83  \\
J0124+0044	&	184	& 65	&  22	  \\
\enddata
\end{deluxetable}

\end{document}